# Quantum Confined Luminescence in Two dimensions

Saiphaneendra Bachu[1], Fatimah Habis[2, 3], Benjamin Huet[1, 4], Steffi Y. Woo[5], Leixin Miao[1], Danielle Reifsnyder Hickey[1, 4, 6], Gwangwoo Kim[7, 8], Nicholas Trainor[1], Kenji Watanabe[9], Takashi Taniguchi[10], Deep Jariwala[7], Joan M. Redwing[1, 4], Yuanxi Wang[2], Mathieu Kociak[5], Luiz H.G. Tizei[5, *], Nasim Alem[1, 4, *]

[1]Department of Materials Science and Engineering, The Pennsylvania State University, University Park, PA 16802, USA

[2]Department of Physics, University of North Texas, Denton, TX 76203, USA

[3]Department of Physics, Jazan University, Jazan, 82817-2820, Saudi Arabia

[4]2D Crystal Consortium (2DCC), Materials Research Institute (MRI), The Pennsylvania State University, University Park, PA 16802, USA

[5]Université Paris-Saclay, CNRS-UMR 8502, Laboratoire de Physique des Solides, Orsay, 91405, France

[6]Department of Chemistry, The Pennsylvania State University, University Park, PA 16802, USA

[7]Department of Electrical and Systems Engineering, University of Pennsylvania, Philadelphia, PA 19104, USA

[8]Department of Engineering Chemistry, Chungbuk National University, Chungbuk 28644, Republic of Korea

[9]Research Center for Electronic and Optical Materials, National Institute for Materials Science, 1-1 Namiki, Tsukuba, 305-0044, Japan

<sup>10</sup>Research Center for Materials Nanoarchitectonics, National Institute for Materials Science, 1-1 Namiki, Tsukuba, 305-0044, Japan

*Corresponding authors: luiz.galvao-tizei@universite-paris-saclay.fr, nua10@psu.edu

**Abstract**

Achieving localized light emission from monolayer two-dimensional (2D) transition metal dichalcogenides (TMDs) embedded in the matrix of another TMD has been theoretically proposed but not experimentally proven. In this study, we used cathodoluminescence performed in a scanning transmission electron microscope to unambiguously resolve localized light emission from 2D monolayer $MoSe_2$ nanodots of varying sizes embedded in monolayer $WSe_2$ matrix. We observed that the light emission strongly depends on the nanodot size wherein the emission is dominated by $MoSe_2$ excitons in dots larger than 85 nm, and by $MoSe_2/WSe_2$ interface excitons below 50 nm. Interestingly, at extremely small dot sizes (< 10 nm), the electron energy levels in the nanodot become quantized, as demonstrated by a striking blue-shift in interface exciton emission, thus inducing quantum confined luminescence. These results establish controllable light emission from spatially confined 2D nanodots, which holds potential to be generalized to other 2D systems towards future nanophotonic applications.

**Introduction**

Localized light emission in nanoscale materials has attracted significant interest and opened a wide new field of nanophotonics towards applications such as lasers,[1] light emitting devices[2] and photodetectors[3] among others. Traditionally, such localized light emission has been achieved by nanoparticles/quantum dots,[4] nanowires,[5] quantum wells,[6] and point defects in semiconductors and wide bandgap crystals.[7] In these material platforms, spatial confinement of the emission centers often results in quantized energy levels with unique light emission characteristics, sometimes even leading to single photon emission (SPE).[8] These characteristics have the potential to advance technologies in quantum emission,[9] quantum communication,[10] and quantum cryptography.[11]

Two-dimensional (2D) transition metal dichalcogenides (TMDs) are an emerging platform to achieve localized light emission owing to their band structure tunability and large exciton binding energies.[12,13] Localized emission has been reported in engineered 2D TMDs using irradiation induced point defects,[14] edges,[15] formation of bubbles,[16] electrostatic gating,[17] and strain patterning with nanopillar supports.[18] However, all these approaches are top-down in nature and hence are difficult to control and scale up. A few recent reports presented an alternative carrier localization technique in TMD wherein one TMD is spatially confined in the form of a nanodot within the matrix of another TMD.[19–21] This class of nanodot/matrix in-plane heterostructures also have the potential to introduce quantum confinement effects which can further modify the light emission from the nanodot area.[19,20] Compared with top-down methods, this approach is suitable for bottom-up synthesis and holds greater prospects in scalability, tunability of dot sizes, and control of the nanodot/matrix interface *via* selection of appropriate TMD pairings. However, realizing fully functional 2D TMD nanodot/matrix heterostructures faces challenges in growth control and nanoscale characterization. First, scalable synthesis of in-plane nanodot/matrix in-plane heterostructures requires fine control over precursor flow and pressure. We recently addressed this challenge using metal-organic chemical vapor deposition (MOCVD) by sequentially supplying the nanodot precursors followed by matrix precursors.[22,23] Second, resolving the emission properties of individual dots requires a technique with spatial resolution comparable to that of nanodot sizes, which is in the order of a few nanometers. Techniques such as photoluminescence (PL) and cathodoluminescence (CL)-in-scanning electron microscope (SEM)[13] lack the required spatial resolution to investigate the effects of heterogeneities in nanodot sizes and shapes and are likely to provide the macroscale luminescence behavior averaged over multiple dots. To overcome this challenge, we performed CL inside an aberration-corrected scanning transmission electron microscope (STEM), offering superior spatial resolution that is more suitable to study the nanodot/matrix heterostructures.[24,25]

This study focuses on understanding the luminescence behavior of $MoSe_2$ nanodots in monolayer $MoSe_2$/$WSe_2$ nanodot/matrix in-plane heterostructure films (**Figure 1a**), encapsulated with hexagonal boron nitride (hBN) flakes, and probed using CL performed in STEM at cryogenic temperatures (**Figure 1b**). We investigated the effects of nanodot size on the light emitting characteristics such as emission energy/wavelength, spatial emission signature, and the carrier confinement. In the large-dot-size limit (> 85 nm), emission energies approach those of $MoSe_2$ monolayers, where all the emission is localized within the dots (**Figure 1g, 1h**). In the intermediate size range (10 – 50 nm), dot emission is dominated by the interface excitons with a ring-like emission signature where the ring size is much larger than the nanodot sizes. Finally, below 10 nm, the dots exhibit quantum confinement effects evidenced by the blue-shift in interface exciton emission energies (**Figure 1e**) with solid hotspot-like emission signature. Our unique experimental approach allows for direct visualization of the atomic and chemical structure of individual nanodots and enables directly linking the atomic structure to their emission characteristics.

**Results and Discussion**

The $MoSe_2$/$WSe_2$ nanodot/matrix heterostructure films with varying dot sizes were synthesized using MOCVD (see **Methods** for a description of the synthesis protocol). The synthesis process results in an almost fully coalesced monolayer film, wherein the $MoSe_2$ nanodots are embedded in the matrix of $WSe_2$, with occasional $WSe_2$ bilayers on top (**Figure S1a**). A high-magnification annular dark-field scanning transmission electron microscopy (ADF-STEM) image confirms the presence of the nanodots in the film owing to the atomic number difference between the Mo and W (**Figure S1b**). Atomic-resolution ADF-STEM imaging reveals that the $MoSe_2$ nanodots exhibit a truncated triangular shape (**Figure 1a** and **Figure S1c**). A closer look at the interface indicates the atomic-level sharpness with minimal alloying (**Figure S1d**) and the corresponding fast Fourier transformation (FFT) pattern (**inset** in **Figure S1d**) suggests the in-plane epitaxy between the nanodot and the matrix. Energy dispersive spectroscopy (STEM-EDS) elemental mapping (**Figure S2**) shows a clear separation between the Mo and W signals, further proving the sharp interface between the nanodot and the matrix. Moreover, a selected area electron diffraction (SAED) pattern obtained from a large area of the heterostructure film (**inset** in **Figure 1a**) illustrates the single-crystalline nature of the film and highlights the oriented growth on the sapphire substrate. It is noteworthy that the reflections of $MoSe_2$ and $WSe_2$ could not be separated in both the SAED and the FFT patterns because of the close lattice matching between the nanodot and the matrix (0.3 % mismatch).[26] In fact, $MoSe_2$ and $WSe_2$ with close lattice matching were chosen here because they would form heterostructures with coherent interfaces, no interfacial defects, and no strain that may induce significant band bending.[26] The absence of defects, monovacancies or lattice mismatch strain along the interface, as seen in the high-resolution STEM image (**Figure S1d**), also ensures that the contribution of defects in the light emission behavior of these in-plane heterostructures is insignificant.

We probed the optical properties of $MoSe_2$/$WSe_2$ heterostructures at nanoscale with the electron beam inside a STEM and collected electron energy loss spectroscopy (EELS) and CL signals, as depicted in **Figure 1b**. To enhance the light emission intensity, the $MoSe_2$/$WSe_2$ heterostructure films were encapsulated with hBN flakes on top and bottom (see **Methods** for detailed information about the sample preparation and EELS/CL data acquisition). Moreover, the hBN encapsulation helps achieving spectra with smaller peak widths and reduce the roughness of the atomically thin layers.[25,27] A typical EELS spectrum (**Figure 1c**) obtained from the heterostructure shows the A exciton peaks of the $MoSe_2$ nanodot ($X_{Mo}$) and the $WSe_2$ matrix ($X_W$) at 1.635 eV and 1.710 eV, respectively. The peaks at 1.849 eV and 2.146 eV correspond to the B excitons of $MoSe_2$ ($X^B_{Mo}$) and $WSe_2$ ($X^B_W$), respectively. The EELS peak positions of A and B excitons agree well with the previously reported values.[28,29] The additional peaks seen at higher energies are indicative of strong absorption due to the band nesting in $MoSe_2$ and $WSe_2$.[29] The summed CL spectrum obtained from an area of the heterostructure spanning a few $MoSe_2$ nanodots of various sizes (**Figure 1d**) shows peaks at 1.639 eV, 1.721 eV and 1.545 eV, corresponding to the $MoSe_2$ A exciton ($X_{Mo}$), $WSe_2$ A exciton ($X_W$) and interface exciton ($X_{Mo/W}$) emissions, respectively.[30] A schematic illustration of these three types of excitons is overlaid on **Figure 1a** showing $X_{Mo}$ and $X_W$ are formed by holes and electrons of $MoSe_2$ and $WSe_2$, respectively, whereas $X_{Mo/W}$ is formed by a hole from $WSe_2$ and an electron from $MoSe_2$. The $X_{Mo}$ and $X_W$ CL peak positions are consistent with their EELS counterparts, within our experimental error (~ 3%, i.e., 50 meV at 1.63 eV, for EELS). We also observe the emergence of a new peak at 1.589 eV ($X^*_{Mo/W}$), originating from blue-shifted interface excitons, which will be discussed in detail later.

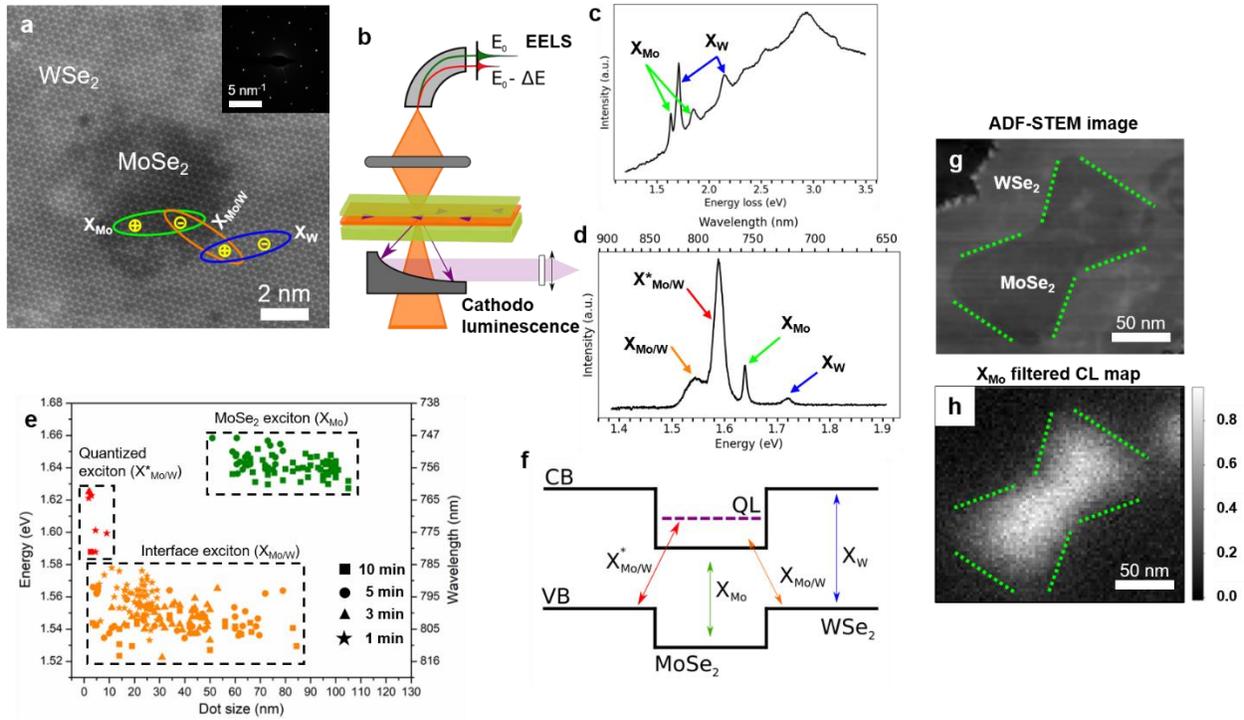

**Figure 1. Morphology of the heterostructures and Effect of MoSe₂ nanodot size on the light emission characteristics.** (a) atomic-resolution ADF-STEM image of MoSe₂/WSe₂ nanodot/matrix in-plane heterostructure; inset shows a SAED pattern collected from a large area, highlighting the single-crystalline nature of the as-grown heterostructure, (b) schematic displaying the EELS- and CL-in-STEM experiment performed on the heterostructure encapsulated with hBN on top and bottom; green layers are the hBN flakes, the orange layer sandwiched between the hBN layers is the TMD layer, and the purple triangles in the TMD layer represent the MoSe₂ dots, (c) EELS spectrum of the 10 min heterostructure sample summer over a region containing one dot showing A and B exciton peaks of MoSe₂ and WSe₂, (d) CL spectrum of the 10 min heterostructure sample summer over an area spanning a few MoSe₂ nanodots of various sizes, showing MoSe₂ exciton (X$_{Mo}$), WSe₂ exciton (X$_W$), interface exciton (X$_{Mo/W}$) and quantized exciton (X*$_{Mo/W}$) emission peaks, (e) scatter plot of CL peak position vs MoSe₂ nanodot size illustrating clusters of three emission modes depending on the dot size (X$_{Mo}$ – green data points, X$_{Mo/W}$ – orange data points and X*$_{Mo/W}$ – red data points); shapes of the data points indicate the sample wherein the squares, circles, triangles and stars belong to 10-, 5-, 3- and 1- min samples, respectively, (f) flat-band energy band diagram illustrating the type-II band alignment between MoSe₂ and WSe₂; QL refers to a quantized electron energy level inside the MoSe₂ nanodot, (g) low-magnification ADF-STEM image of the MoSe₂/WSe₂ nanodot/matrix in-plane heterostructure showing two large MoSe₂ nanodots merged in the shape of a bow-tie; the top-left corner of the image shows an uncoalesced edge of the heterostructure, and (h) corresponding normalized filtered CL map of X$_{Mo}$ peak showing the emission is completely localized within the MoSe₂ nanodots.

Next, we studied the effect of the MoSe₂ dot size on the light emission characteristics and the possibility of quantum confinement effects at extremely small sizes. The variation in the dot size is achieved by changing the nucleation time of $MoSe_2$ nanodots from 1 minute to 3, 5 and 10 minutes during the heterostructure synthesis (see **Methods**). ADF-STEM images obtained from the four samples (**Figure S3a–d**) confirm that the dot size increases with the nucleation time during the MOCVD growth (see **Figure S3e** for nanodot size measurement procedure). Next, the positions of the strongest CL emission peak originating from 240 individual $MoSe_2$ nanodots are plotted against their dot sizes across all the four samples, as shown in **Figure 1e** (see **Methods** for the details about how each data point in the plot is generated). In the scatter plot, the data points are colored based on the dominant emission mode exhibited by each dot ($X_{Mo}$ is green, $X_{Mo/W}$ is orange and $X^*_{Mo/W}$ is red).

It is apparent from the scatter plot that the data points of different emission modes cluster into different nanodot size groups across all the samples. First, large $MoSe_2$ nanodots ($> 85$ nm) always show the $X_{Mo}$ peak as the strongest peak whereas the dots with sizes in the range of $50 - 85$ nm exhibit either the $X_{Mo}$ or the $X_{Mo/W}$ peak as the strongest peak. The average emission wavelength of the $X_{Mo}$ peak is 1.642 eV and all the dots exhibiting this peak belong to either the 5- or 10- min samples. Interestingly, the average wavelength of the $X_{Mo}$ peaks observed in the 5 min sample (1.646 eV) is slightly blue shifted compared to the peaks observed in the 10 min sample (1.640 eV), matching the PL observation from these samples, as shown in **Figure S4** (also, see this Ref.[23] for a detailed optical study). Nanodots in the size range of $10 - 50$ nm mainly show the $X_{Mo/W}$ peak, which is a significant result because the existence of interface excitons in 2D in-plane heterostructures has long been predicted in literature but only observed experimentally in one report.[31] We attribute the emission from $X_{Mo/W}$ and the suppression of $X_{Mo}$ emission in this size range to the increase in the interface exciton binding energy with decrease in dot size.[31,32] Moreover, for smaller dot sizes ($< 50$ nm), the interface exciton size becomes comparable to the dot size, leading to higher binding energies than interface excitons at an ideal (infinitely long) interface.[32] The average emission energy of the $X_{Mo/W}$ peak from the data points of all the four samples is 1.551 eV. The relative red shift of the $X_{Mo/W}$ peak position with respect to $X_{Mo}$ can be understood by referring to the band alignment at the heterostructure interface (**Figure 1f**). Semiconducting $MoSe_2$ and $WSe_2$ form type-II band alignment with two staggered band gaps resulting in an interfacial band gap that is smaller than that of both $MoSe_2$ and $WSe_2$.[33] The numerical value of the difference in the peak positions of $X_{Mo}$ and $X_{Mo/W}$ (1.642 – 1.551 = 0.091 eV) agrees well with the value probed by the optical measurements performed in the sole experimental study till date.[31] Moreover, a redshift of this order cannot be due to trions, since the trions in $MoSe_2$ are red shifted by only 0.03 eV even under heavily gated conditions.[34]

Most strikingly, for dot sizes below 10 nm, we observe a prominent blue-shift in the $X_{Mo/W}$ emission. We attribute this to the quantization of the electron energy levels in the $MoSe_2$ nanodots, as predicted in ref.,[19] and hence call it a quantized exciton ($X^*_{Mo/W}$) emission (see **Figure 1f** for the flat-band energy diagram and **Figure 5** for more details). The average wavelength of the $X^*_{Mo/W}$ emission is 1.607 eV, positioned in between the $X_{Mo}$ and $X_{Mo/W}$ peaks. Representative spectra from each unique subset of datapoints of the three clusters are displayed in **Figure S5**, which clearly illustrates the typical evolution of CL spectrum with decreasing $MoSe_2$ nanodot size, as explained so far. The distribution of $X_W$, $X_{Mo}$, $X_{Mo/W}$ and $X^*_{Mo/W}$ peak positions are also displayed as histograms plotted for each of the four samples in **Figure S6**. We also mapped the spatial light emission signature of various excitons observed in the heterostructure system. For example, **Figures 1g-h** depict the emission map of $MoSe_2$ exciton ($X_{Mo}$) that is dominant in the dots larger than 85 nm, which shows the light emission is completely localized within the $MoSe_2$ nanodots. Moreover, the size and shape of the emission hotspot closely match that of the merged nanodots. To the best of our knowledge, such localized light emission from nanodot/matrix in-plane heterostructures have never been reported so far. This represents a major breakthrough in the field of nanophotonics because it opens a way to engineer and directly visualize spatially confined 2D nanodots. EELS spectra of 1-, 5- and 10- min samples (**Figure S7**) show increasing intensity of the $X_{Mo}$ peak relative to that of $X_W$, indicative of increasing area fraction of MoSe$_2$ nanodots with increasing nucleation time.

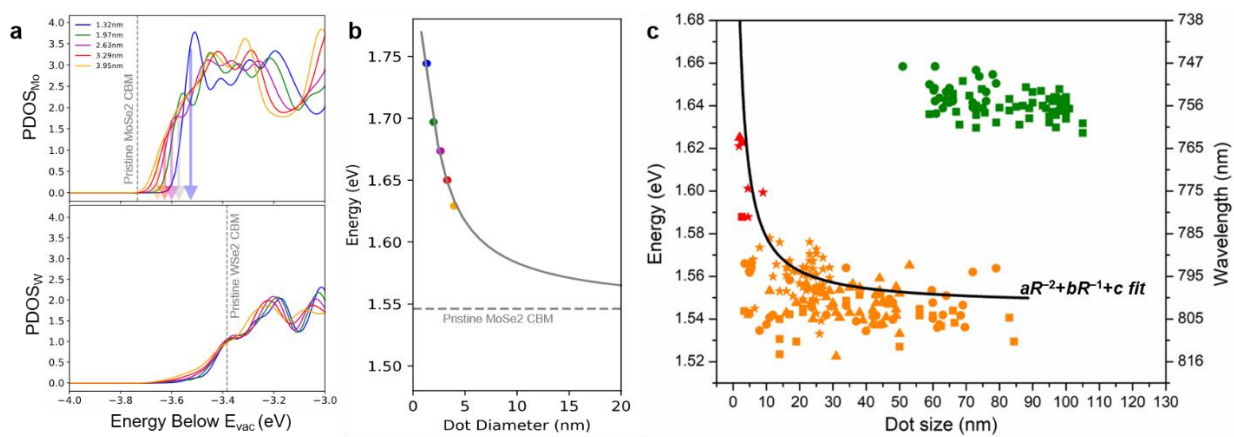

**Figure 2. Simulation of the effect of nanodot size on the electronic band structure of MoSe$_2$ and the interface exciton ($X_{Mo/W}$) emission.** (a) density of states (DOS) projected onto Mo (upper) and W (lower) for five MoSe$_2$ nanodot sizes, 3.95 (orange), 3.29 (red), 2.63 (purple), 1.97 (green) and to 1.32 nm (blue); for the Mo-projected DOS, the conduction band edge is seen to increase with smaller dot sizes, (b) the conduction band edge energies of the nanodots modeled, along with their asymptote towards pristine MoSe$_2$ conduction band edge, can be fitted to extrapolate how nanodot bands vary with size, and (c) predicted variation of $X_{Mo/w}$ peak position with respect to the MoSe$_2$ nanodot size overlaid onto the scatter plot from **Figure 1e**.

To theoretically capture the measured relation between emission energy and dot size, we performed first-principles calculations using atomic models of nanodot/matrix heterostructures and extrapolated towards the large dot size limit (see **Methods** for more details). We considered five hexagonal MoSe$_2$ nanodots with diameters 1.32, 1.97, 2.63, 3.29 and 3.95 nm and calculated their respective density of states (DOS) projected onto Mo and W, as shown in **Figure 2a**. All displayed energies are relative to the vacuum potential. We observe that the onset of MoSe$_2$ conduction band DOS moves up away from the pristine $MoSe_2$ band edge as the dot size decreases, whereas the conduction band edge of the $WSe_2$ matrix remains largely unchanged. The trend for $MoSe_2$ conduction band edge against dot diameter is plotted separately in **Figure 2b** and fitted with $aD^{-2}+bD^{-1}+c$,[19] where $D$ is dot diameter, to extrapolate the amount of blue shift in the conduction band minima (CBM) for any given dot size. Next, we used this trend to plot the expected dependence of interface exciton emission ($X_{Mo/W}$) peak position against the dot size by setting the experimental average value of $X_{Mo/W}$ peak as its asymptotic value. The final scatter plot combining theoretical and measured emission energies in **Figure 2c** shows good agreement.

The spatial signatures of the three light emission modes discussed so far ($X_{Mo}$, $X_{Mo/W}$ and $X^*_{Mo/W}$) are detailed in **Figures 3–5**. We start by presenting the spatial signature of the $X_{Mo}$ light emission in **Figure 3**. A large field-of-view ADF-STEM image of the heterostructure encapsulated in hBN on a Quantifoil support is shown in **Figure 3a**. A small area outlined by the red rectangle containing two $MoSe_2$ nanodots (**Figure 3b**) is considered for the CL spectrum imaging (see **Methods**). While both the dots in this area are present in the monolayer of the film, the bottom nanodot (dotted red circle) is covered with a second layer of $WSe_2$ on top (the edge of the second layer is highlighted by dashed yellow line). The summed CL spectrum obtained from this area (**Figure 3c**) shows two peaks at 1.631 eV and 1.771 eV corresponding to $X_{Mo}$ and $X_W$ emissions, respectively. As expected, we do not observe any emission from the interface or quantized excitons here because of the large sizes of the dots in this region (> 85 nm). We believe the sloping background in the spectrum at higher energies/lower wavelengths could be attributed to the light emission from the hBN layers used for encapsulation[35] (see **Figure S8** for further evidence). Filtered map acquired by placing a window around the $X_{Mo}$ emission peak (**Figure 3d**) reveals the localized light emission from within the top nanodot (dashed green triangle), similar to the dots shown in **Figure 1g**. The higher intensity of the $X_{Mo}$ peak compared to the $X_W$ peak in the CL spectrum, despite the smaller area fraction of the MoSe$_2$ nanodots, could be attributed to the higher quantum efficiency of localized excitons.[36] Strikingly, the bottom nanodot underneath the second layer of WSe$_2$ does not appear to emit any light (dotted red circle). This could be attributed to the ultrafast charge transfer between the MoSe$_2$ and WSe$_2$ layers in the locally formed WSe$_2$/MoSe$_2$ vertical heterostructure, which prevents the formation of $X_{Mo}$ and consequently quenches the light emission.[37] Filtered map of the $X_W$ peak showing light emission from around the top MoSe$_2$ nanodot in the monolayer area is presented in **Figure S9**.

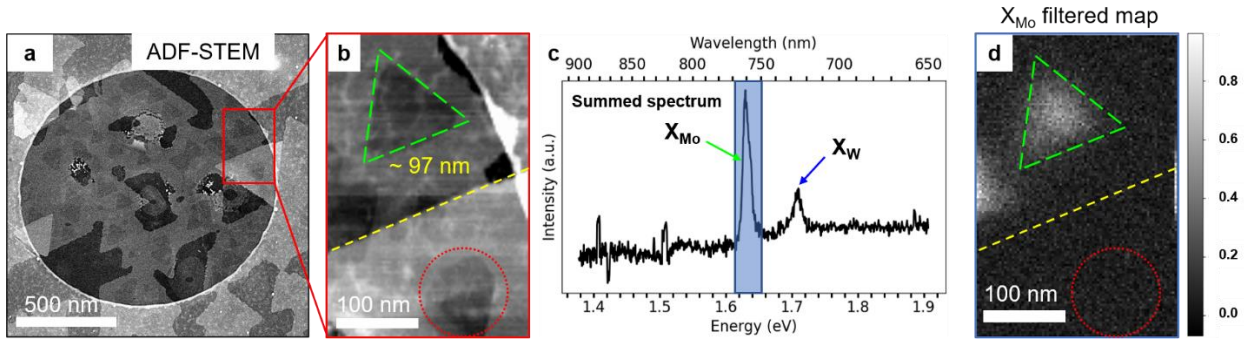

**Figure 3. Emission map of $X_{Mo}$ observed in large dots.** (a) low magnification ADF-STEM image of heterostructure sandwiched between hBN flakes on a Quantifoil support, (b) enlarged ADF-STEM image of the area outlined by the red rectangle in (a) showing two MoSe$_2$ nanodots; the nanodot on the bottom is covered with a second layer of WSe$_2$ on top; contrast between the nanodots and the matrix is blurred by the presence of hBN on top and bottom, and interfacial/surface carbon webbing present in some areas, (c) summed CL spectrum obtained from the area shown in (b), and (d) filtered map of $X_{Mo}$ peak showing that the emission is localized within the top nanodot; the bottom nanodot does not emit any light as it is covered by a second layer of WSe$_2$ on top.

Interface exciton ($X_{Mo/W}$) emission from the heterostructure is similarly mapped and presented in **Figure 4**. **Figure 4a** shows a MoSe$_2$ nanodot with 11 nm size and the summed CL spectrum obtained from this area (**Figure 4b**) displays two peaks at 1.540 eV and 1.722 eV corresponding to $X_{Mo/W}$ and $X_W$ emissions, respectively. The $X_{Mo}$ and $X*_{Mo/W}$ peaks are not seen in the spectrum because of the absence of both large dots (> 50 nm) and extremely small dots (< 10 nm) in this area of the heterostructure film. The filtered map of the $X_{Mo/W}$ emission (**Figure 4c**) shows a ring-like pattern with a bright spot at the center. The location of the central bright spot falls on top of the 11 nm nanodot suggesting that the emission originates from that particular dot. The bright pixels forming the ring indicate the other positions where carrier excitations are created by the scanning electron beam which eventually diffuse to the small $MoSe_2$ dot where excitons radiatively recombine across the interface to emit light.[24] Having a delocalized light emission in cathodoluminescence is well-known,[24] but ring-like features have not been reported yet to the best of our knowledge. However, we can infer that the emission sites are the $MoSe_2/WSe_2$ interfaces because the emission energy/wavelength matches that of interface exciton emission energy/wavelength.[31] The diameter of the emission ring is in the range of 120 – 140 nm suggesting that the carriers excited by the electron beam (in the encapsulating hBN), located at 60 – 70 nm away from the nanodot diffuse to the nanodot/matrix interface to form excitons, recombine and emit light. This diffusion length values agrees well with the reported typical excitation diffusion distances in hBN.[38]

When the outline of the emission ring is overlaid on top of **Figure 4a** (dashed white outline), it is apparent that the signal along the ring quenches where it meets an uncoalesced free edge of the TMD monolayer film (on the left side) and at the edge of the second layer of $WSe_2$ (dashed cyan line on the top right side). Furthermore, the horizontal dark streaks in the ring pattern could be attributed to the well-known blinking behavior of quantum dots (QD) caused by charging of the dots, wherein the dots alternate between on-state and off-state.[39] Following a common mechanism of QD blinking in core/shell QD structures,[39] we presume that the on-state is associated with a charge-neutral dot where excitons can radiatively recombine. On the other hand, the off-state is associated with a charged dot (state initiated by Auger ionization and terminated upon neutralization) where exciton recombination energies are non-radiatively transferred to the excess charge through Auger recombination (**Figure 4d**). Another interesting observation from the CL map is the negligible emission detected between the central spot and the outer ring, which we attribute to the same charging behavior of quantum dots that also causes the streaks in the ring. Since the dark interior of the ring has similar intensity as the dark streaks, we hypothesize the dark interior is caused by the dot persisting in a charged off-state. This is also supported by the well-established photodarkening behavior of QDs,[40] where a higher laser power increases the rates at which dots are populated by two electron-hole pairs (a prerequisite of Auger ionization) and hence increases overall ionization rates. Thus, we conclude that the high-CL-intensity rings are a product of the electron beam close enough to the small dots to allow carrier diffusion but not too close to photodarken the dots. We note that this is one hypothesis to explain the experimentally observed ring pattern of the interface exciton emission among other possible mechanisms that cannot be ruled out. For illustration, we used a minimal 2D random walk model parametrized by the observed ring (**Figure 4e**). To simulate a STEM-CL scan region, we performed random walk for excited carriers with starting positions at each 4 nm × 4 nm pixel within a square region of > 100 nm width. All carriers performing random walk are assigned a finite lifetime that achieves a diffusion length of 50 nm. We then monitored the percentage of carriers arriving at a small (10 nm) nanodot placed at the origin (green triangle). A pixel is colored in grayscale ranging from 0% of the carriers created at the pixel arriving at the dot (black) to 1% arriving (white). When the dot accepts more than 1% of total carriers created, it is considered charged and photodarkened and thus the corresponding pixel (where carriers were created) is then left to be black. The ring shape shows that only carriers initiated at a radius of ~ 50 nm away from the nanodot have sufficient chance of reaching but not overpopulating the nanodot.

While this example shows one isolated ring pattern, we noticed multiple ring-like emission patterns overlapping with each other originating from multiple nanodots in close proximity when we performed CL over large areas (two such example $X_{Mo/W}$ emission maps are in **Figure S10**). For each ring observed in these maps, we discovered a nanodot at the location of the central spot. Notably, we observe that the sizes of the rings roughly remain constant despite the change in nanodot size. This observation also supports the explanation of ring patterns due to the diffusion of the carrier excitations.[24,30,38] Occasionally, we observed ring shape distorting, for example, into a heart shape, as seen inside the dashed red box in **Figure S10b**. To explain this observation, we extended the aforementioned 2D random walk model by bringing a second dot, labeled by a red circle, close to the original nanodot (**Figure S10c**). The second dot acts as a sink drawing away much of the carrier population created by the CL excitation, leading to a finite intensity closer to the original dot, and thus warps the ring into a heart shape. It should be noted that this is just one possible way a neighboring dot can affect the ring pattern. But, the investigation of the exact nature of the interactions between the dots is complicated and outside the scope of this manuscript.

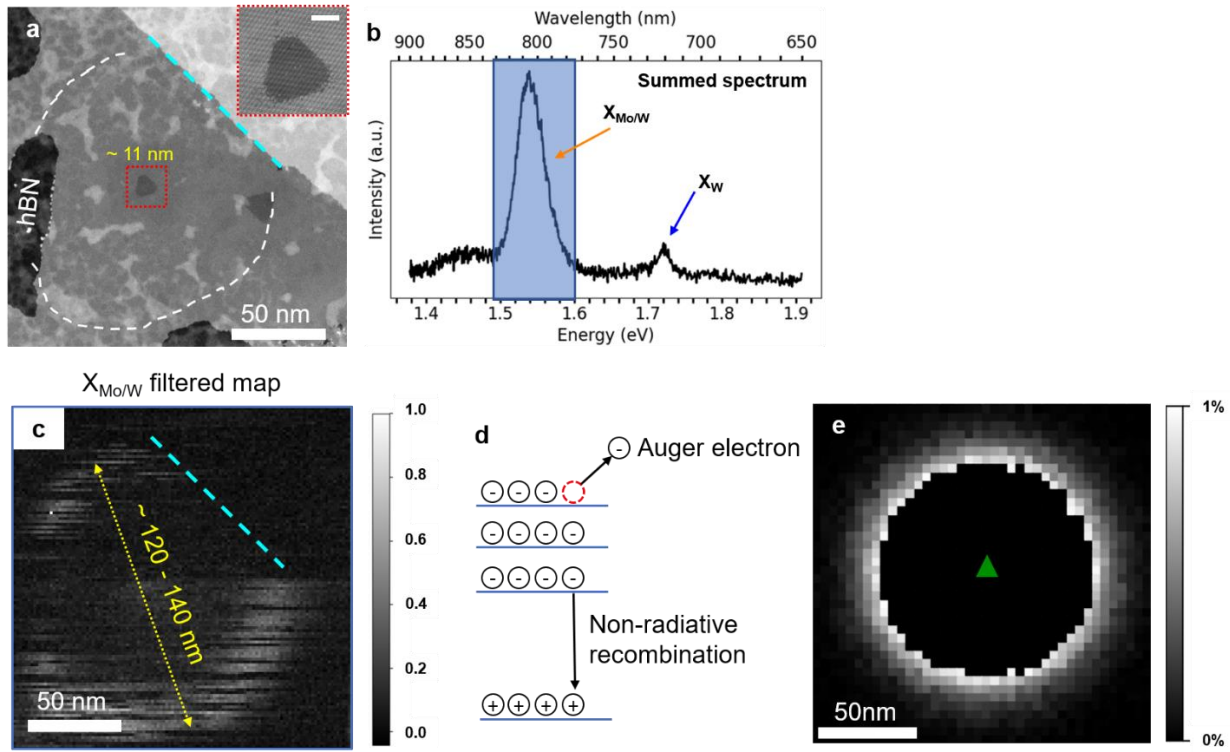

**Figure 4. Emission map of X$_{Mo/W}$ observed in intermediate sized dots.** (a) ADF-STEM image of heterostructure showing a 11 nm sized MoSe$_2$ nanodot (dashed red square); the inset shows a magnified image of the 11 nm dot, (b) summed CL spectrum obtained from the area shown in (a), (c) filtered map of X$_{Mo/W}$ peak showing a ring-like emission pattern with a bright central spot, where the outline of the ring lies far away from the nanodot/matrix interface; the ring is interrupted by the presence of the uncoalesced edge of the heterostructure on the left and the edge of the second WSe$_2$ layer on top right side (dashed cyan line); the outline of ring pattern is overlaid on panel (a) as dashed white line, (d) schematic illustrating the Auger ionization processing leading to charging and photodarkening of MoSe$_2$ nanodots, and (e) demonstration of the ring pattern formation using a minimal 2D random walk carrier diffusion model; green triangle represents a small MoSe$_2$ triangle (~ 10 nm). Scale bar in the inset in panel (a) is 5 nm.

Besides the emission modes discussed so far, we also investigated the emission signature of quantized excitons (X*$_{Mo/W}$) dominant in nanodots below 10 nm size (**Figure 5**). The area of the heterostructure film shown in **Figure 5a** contains three MoSe$_2$ nanodots with sizes from the three different size ranges of **Figure 1e** (the three dots are highlighted with dashed outlines in **Figure S11a**, a duplicate of **Figure 5a**). As a result, the summed CL spectrum obtained from this region shows four distinct emission peaks, namely X$_{Mo/W}$, X*$_{Mo/W}$, X$_{Mo}$ and X$_W$ (**Figure 5b**). The emission maps of X$_{Mo/W}$, X$_{Mo}$ and X$_W$ are presented in **Figure S11c-e**. The quantized exciton (X*$_{Mo/W}$) peak at 1.591 eV in **Figure 5b** is attributed to the small dot seen in this area (highlighted by dotted red square and enlarged as **Figure 5a** inset). The spatial signature of this emission (**Figure 5c**) shows a solid hotspot of 160 – 190 nm in size, that is much larger than the dot itself (the outline of the emission hotspot is overlaid on **Figure 5a** as a dashed white line for reference). Interestingly, the small dot seen here does not exhibit a triangular or a hexagonal shape, rather it looks almost circular like a nucleus that has just been formed. For dots with this shape, we estimated their size as the diameter of the largest circle that completely fits the nanodot, which turns out to be 2.7 nm for this dot. It is noteworthy that among the datasets we analyzed to generate the scatter plot in **Figure 1e**, whenever we observed the X*$_{Mo/W}$ emission peak, a small dot (< 10 nm) was discovered in that area exhibiting a large emission hotspot. The spatial and spectral separation of the emission signals from the X$_{Mo/W}$ and X*$_{Mo/W}$ peaks is further confirmed by using unsupervised machine learning based dimensionality reduction and clustering[41] (see **Methods** and **Figure S12**). Overall, these results unambiguously confirm the quantum confinement of excitons in extremely small monolayer MoSe$_2$ nanodots with unique emission signature and it was possible to spatially and spectrally resolve them owing to the superior resolution capabilities of cathodoluminescence performed inside a STEM.

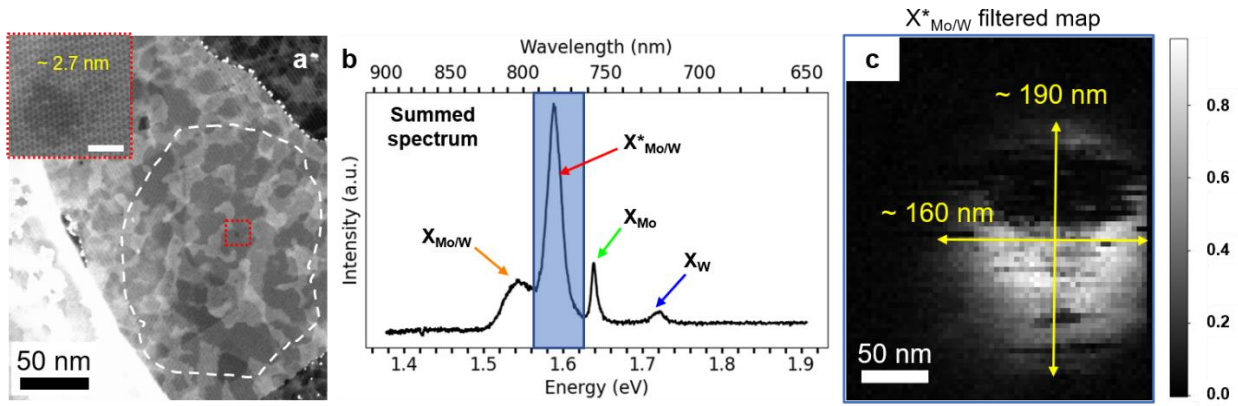

**Figure 5. Emission map of X\*$_{Mo/W}$ observed in small dots.** (a) ADF-STEM image of heterostructure showing MoSe$_2$ nanodots with different sizes (all the triangles are highlighted in **Figure S11a**); contrast between the nanodots and the matrix is blurred by the presence of hBN on top and bottom, and interfacial/surface carbon webbing present in some areas; the smallest dot of ~ 2.7 nm size is outlined with dotted red box and enlarged as the inset, (b) summed CL spectrum obtained from the area shown in (a), and (c) filtered map of X\*$_{Mo/W}$ peak originating from the 2.7 nm dot, showing an emission hotspot that is much larger than the nanodot; the outline of the emission hotspot is overlaid on panel (a) as a dashed white line for reference. Scale bar in the inset in panel (a) is 2 nm.

**Conclusions**

In summary, this study elucidates the emission characteristics from the monolayer TMD nanodots and directly links them to their individual atomic structure. We have experimentally shown that it is possible to achieve localized light emission from laterally confined MoSe$_2$ nanodots embedded in the WSe$_2$ matrix. In addition, we were able to observe the competing luminescence behavior between multiple excitons with varying spatial signatures as the nanodot size is changed, ultimately leading to the quantization effects in the smallest nanodot sizes. Overall, the findings of this study lay the foundation to probe the spatial heterogeneities in the optical properties of nanodots in the 2D limit. This is especially important because 2D materials are the thinnest crystals currently relevant in electronics and optoelectronics. This study shows the promise of the application of 2D heterostructures with nanodot sizes as small as 2 nm leading to significant light emission, making them prominent candidates to realize quantum technologies.

**Methods**

*Synthesis of MoSe$_2$/WSe$_2$ nanodot/matrix in-plane heterostructures*

MoSe$_2$/WSe$_2$ nanodot/matrix in-plane heterostructure films were synthesized over 2-inch c-plane sapphire substrates using a 3-step process in a horizontal cold-wall MOCVD reactor. The three steps involved in the synthesis process are namely, nucleation, consolidation and growth. During the nucleation step, H$_2$Se and Mo(CO)$_6$ gases are supplied to deposit the MoSe$_2$ nanodots on the substrate. The Mo precursor supply is stopped during the consolidation step to facilitate the diffusion of precursor species on the substrate and ensure the formation of the dots with sharp edges. Next, the growth step is initiated by introducing the flow of $W(CO)_6$, which enables the $WSe_2$ matrix to grow around the $MoSe_2$ nanodots and coalesce, leading to the formation of $MoSe_2/WSe_2$ nanodot/matrix in-plane heterostructure. To begin the synthesis procedure, metal precursors ($Mo(CO)_6$ and $W(CO)_6$) were housed in stainless steel bubblers maintained at 950 Torr and 10 °C and the precursor vapors were transported to the reactor via a hydrogen carrier gas. The substrate was initially heated in $H_2$ to 900 °C at a reactor pressure of 200 Torr and annealed for 10 minutes. The nucleation of the $MoSe_2$ QDs was initiated by simultaneously introducing $Mo(CO)_6$ at a flow rate of $6.1 \times 10^{-3}$ sccm and $H_2Se$ at 75 sccm into the reactor. The desired size of the $MoSe_2$ nanodots was controlled by adjusting the duration of nucleation step. In this study, we investigated four different samples, with 1, 3, 5 and 10 minutes of nucleation time. The Mo precursor was then switched out of the reactor and the sample was annealed under a flow of $H_2Se$ for approximately 15 minutes during the consolidation step. For the subsequent growth step, the W precursor was then introduced at a flow rate of $8.7 \times 10^{-4}$ sccm for 15 – 20 minutes. The growth step duration was 20 minutes for 1-, 3- and 10- min samples whereas it was 16 minutes for 5 min sample. The shorter growth duration of 5 min sample likely resulted in a slightly less coalesced film compared to others. Finally, the W precursor was switched out of the reactor and the sample was cooled to 300 °C in a mix of $H_2/H_2Se$ environment and was then cooled further to room temperature under $N_2$ flow. Detailed growth and characterization data associated with the samples produced in this study is available at https://m4-2dcc.vmhost.psu.edu/list/data/mfccmO0IwTzT. This includes substrate preparation and recipe data for samples grown by MOCVD in the 2DCC-MIP facility and standard characterization data including AFM images. Additional datasets are available from the corresponding author upon request.

### *S/TEM characterization: Imaging, SAED, EELS and CL*

SAED, general ADF-STEM imaging, and STEM-EDS elemental mapping of nanodot/matrix in-plane heterostructures were performed on freestanding films on TEM grids. Freestanding TEM samples were prepared using the NaOH and PMMA assisted wet transfer method.[42] EELS, CL and ADF-STEM imaging (in the exact same areas where the spectra are collected) were performed on heterostructure films that are encapsulated with exfoliated hBN flakes on top and bottom before being transferred onto TEM grids. We used a modified dry viscoelastic stamping method[27] to prepare the encapsulated stacks where the polymer stamp used is made up of polydimethylsiloxane (PDMS) and polycarbonate (PC). Briefly, a few hBN flakes are first exfoliated onto a Si/SiO$_2$ substrate and one of the flakes is picked up by the polymer stamp with the help of a combination of temperature (~ 120 °C) and pressure. Then, this step is repeated for the as-grown heterostructure film followed by another hBN flake to assemble the hBN/TMD/hBN stack on the polymer stamp. Finally, the assembled hBN/TMD/hBN stack is dropped onto a Au Quantifoil TEM grid and subsequently placed in a chloroform bath for 10 min to remove any leftover polymer residue.

SAED patterns from freestanding films were acquired using a Thermo Fisher TalosX S/TEM microscope operated at an accelerating voltage of 80 kV. STEM-EDS elemental mapping and ADF-STEM imaging were performed on a dual-corrected Thermo Fisher Titan[3] G2 S/TEM microscope operated at an accelerating voltage of 80 kV and a beam semi-convergence angle of 24.2 mrad. The image acquisition parameters used for the freestanding samples are 115 mm camera length, 50 pA screen current and 4 µs dwell time whereas they are 73 mm or 91 mm camera length, 120 pA screen current and 8 µs dwell time for the encapsulated samples. Monochromated EELS and CL were carried out in a modified Nion HERMES 200 (ChromaTEM) operated at an accelerating voltage of 60 kV with the samples held in a liquid nitrogen holder (HennyZ) at around 150 K temperature. The microscope is equipped with a Mönch CL system from Attolight, capable of an energy resolution of 4 meV (about 1 nm at 600 nm wavelength). The EELS and CL datasets were acquired as spectrum images (SPIMs) wherein the focused electron beam of STEM is scanned across a region of the sample at a preset pixel sampling and a spectrum is collected at every scan point. It should be noted that, on the encapsulated samples, the CL and EELS experiments were conducted first before the ADF-STEM imaging so the samples are free from the beam damage induced from the imaging while conducting the spectroscopic measurements.

All the EELS and CL data presented in this work are processed using a combination of Gatan Digital Micrograph and python package Hyperspy.[43] All the spectra displayed in the manuscript are summed spectra over a given region of interest (ROI) in the acquired SPIMs. Before extracting the summed spectrum, a typical EELS dataset is gain corrected, and the zero-loss peak (ZLP) is aligned and subtracted. Whereas for a typical CL dataset, the readout noise is subtracted from the dataset and spike artifacts are removed. The spatial maps of different CL peaks over the acquired SPIMs are extracted by using wavelength-based peak filtering.

### *Variation in the CL peak positions with the nanodot size*

To generate the histograms in **Figure S6a–d**, we collected 65–70 unique ROI summed spectra from each of the four samples and fit the peaks in the spectra with Lorentzian profiles. The fitted peak positions are then utilized to create the histograms. Here, the positions and sizes of the ROI boxes were chosen based on the intensity variations in the CL SPIMs to eliminate the bias from the microstructural features such as local dot size variations and the presence of uncoalesced and second layer edges. A minimum intensity threshold of $3\times10^5$ counts was used to filter out the low-intensity peaks before generating the histograms. To generate the scatter plot in **Figure 1e**, we placed the ROIs to correlate the location of the nanodots and the corresponding light emission mode and signatures ($X_{Mo}$ vs $X_{Mo/W}$ vs $X*_{Mo/W}$). Here, the peak positions are manually recorded as opposed to the fitted peak positions. If a dot shows more than one type of light emission peak (for example, $X_{Mo}$ and $X_{Mo/W}$), only the most intense peak is considered. The process of ascribing the peak to a given dot when there are multiple dots nearby each other is conducted by considering the limits of exciton diffusion and the subsequent learnings from individual dots shown in **Figures 3–5**. The jupyter notebooks used for these analyses can be found here.[44] We did not plot the peak positions of WSe$_2$ excitons in **Figure 1e** because they remain largely unchanged with variation in the MoSe$_2$ nanodot sizes across the four samples, as seen in the histograms in **Figure S6**. The sizes of the dots, whose ROI spectra are considered for the scatter plot, are measured from their high-resolution ADF-STEM images.

*Photoluminescence*

PL spectra shown in **Figure S4** were obtained by using a continuous 633 nm laser with an excitation power of 20 µW and a 50x lens with 0.35 NA.

*Unsupervised machine learning dimensionality reduction and clustering*

To confirm the spatial and spectral separation between the $X_{Mo/W}$ and $X*_{Mo/W}$ peaks, we employed non-negative matrix factorization (NMF) algorithm for the dimensionality reduction, together with, gaussian mixture model (GMM) algorithm for clustering. We implemented these algorithms using in-house developed MATLAB scripts, similar to the previous report.[41] Comparison between the NMF+GMM analysis and the wavelength filtering applied on one of the CL SPIM datasets is presented in **Figure S12**.

*First-principles calculations*

All density functional theory calculations were performed using the Vienna Simulation Package (VASP).[45,46] We used the Perdew-Burke-Ernzerhof (PBE) parametrization of the generalized gradient approximation exchange-correlation functional.[47] Van der Waals corrections for total energies and forces were accounted for using the semiempirical DFT-D3 correction scheme.[48] All calculations employed projector augmented wave pseudopotentials[49,50] and a plane-wave expansion energy cutoff of 500 eV, a relaxation force threshold of 10 meV/Å, and a $k$-point sampling equivalent to $12 \times 12 \times 1$ in a $MoSe_2$ or $WSe_2$ unit cell. All nanodot models are constructed as $MoSe_2$ hexagonal clusters confined within a $WSe_2$ matrix with matching lattice constants. Conduction band edge energies in **Figure 2b** are directly extracted from Kohn-Sham levels, not from peak features in **Figure 2a**. Before comparing the theoretical and experimental values, as shown in in **Figure 2c**, the x-axis of the theoretical plot was adjusted to convert the hexagonal dot diameters to match the convention used for experimentally measuring dot sizes. Here, the measured sizes of experimentally observed truncated triangular dots can be considered equivalent to twice the apothem of a hexagon so a factor of $\sqrt{3}/2$ was used to convert the diameter of a hexagon to 2*apothem of a hexagon.

## Acknowledgements

The authors would like to acknowledge the financial support from National Science Foundation (NSF) through CAREER DMR-1654107, CAREER DMR-2340733, and Penn State 2D Crystal Consortium – Materials Innovation Platform (2DCC – MIP) under DMR-1539916 and DMR-2039351. We gratefully acknowledge the Materials Characterization Lab (MCL) platforms of Penn State University for the characterization equipment. N. Alem also acknowledges the Fulbright Scholar program grant she received through the Franco-American Fulbright

Commission. D. Reifsnyder Hickey also acknowledges start-up funds from the Penn State Eberly College of Science, Department of Chemistry, and Materials Research Institute. N. Trainor acknowledges the National Science Foundation Graduate Research Fellowship Program under Grant No. DGE1255832H. This work has received support from the National Agency for Research under the program of future investment TEMPOS-CHROMATEM (reference no. ANR-10-EQPX-50), the JCJC grant SpinE (reference no. ANR-20-CE42-0020), and from the European Union's Horizon 2020 Research and Innovation Program under grant agreements 823717 (ESTEEM3) and 101017720 (EBEAM). D. Jariwala and G. Kim acknowledge primary support for this work by the Asian Office of Aerospace Research and Development (AOARD) of the Air Force Office of Scientific Research (AFOSR) FA2386-20-1-4074 and partial support from FA2386-21-1-406. K. Watanabe and T. Taniguchi acknowledge support from the JSPS KAKENHI (Grant Numbers 20H00354, 21H05233 and 23H02052) and World Premier International Research Center Initiative (WPI), MEXT, Japan. F. Habis and Y. Wang acknowledge computational resources from the Texas Advanced Computing Center. Part of the modeling work was supported by computational resources from a user project at the Center for Nanophase Materials Sciences (CNMS), a US Department of Energy, Office of Science User Facility at Oak Ridge National Laboratory, and also partially by user project R0076 at the 2DCC-MIP.

**Supporting Information**

Additional STEM, EELS and CL characterization of $MoSe_2$/$WSe_2$ nanodot/matrix in-plane heterostructures

**Author Contributions**

The manuscript was written through contributions of all authors. All authors have given approval to the final version of the manuscript.

**Notes:** The authors declare no competing financial interest.

# Supporting Information for

# Quantum Confined Luminescence in Two dimensions

*Saiphaneendra Bachu[1], Fatimah Habis[2, 3], Benjamin Huet[1, 4], Steffi Y. Woo[5], Leixin Miao[1], Danielle Reifsnyder Hickey[1, 4, 6], Gwangwoo Kim[7, 8], Nicholas Trainor[1], Kenji Watanabe[9], Takashi Taniguchi[10], Deep Jariwala[7], Joan M. Redwing[1, 4], Yuanxi Wang[2], Mathieu Kociak[5], Luiz H.G. Tizei[5, \*], Nasim Alem[1, 4, \*]*

[1]Department of Materials Science and Engineering, The Pennsylvania State University, University Park, PA 16802, USA

[2]Department of Physics, University of North Texas, Denton, TX 76203, USA

[3]Department of Physics, Jazan University, Jazan, 82817-2820, Saudi Arabia

[4]2D Crystal Consortium (2DCC), Materials Research Institute (MRI), The Pennsylvania State University, University Park, PA 16802, USA

[5]Université Paris-Saclay, CNRS-UMR 8502, Laboratoire de Physique des Solides, Orsay, 91405, France

[6]Department of Chemistry, The Pennsylvania State University, University Park, PA 16802, USA

[7]Department of Electrical and Systems Engineering, University of Pennsylvania, Philadelphia, PA 19104, USA

[8]Department of Engineering Chemistry, Chungbuk National University, Chungbuk 28644, Republic of Korea

[9]Research Center for Electronic and Optical Materials, National Institute for Materials Science,

1-1 Namiki, Tsukuba, 305-0044, Japan

[10]Research Center for Materials Nanoarchitectonics, National Institute for Materials Science, 1-1

Namiki, Tsukuba, 305-0044, Japan

*Corresponding authors: luiz.galvao-tizei@universite-paris-saclay.fr, nua10@psu.edu

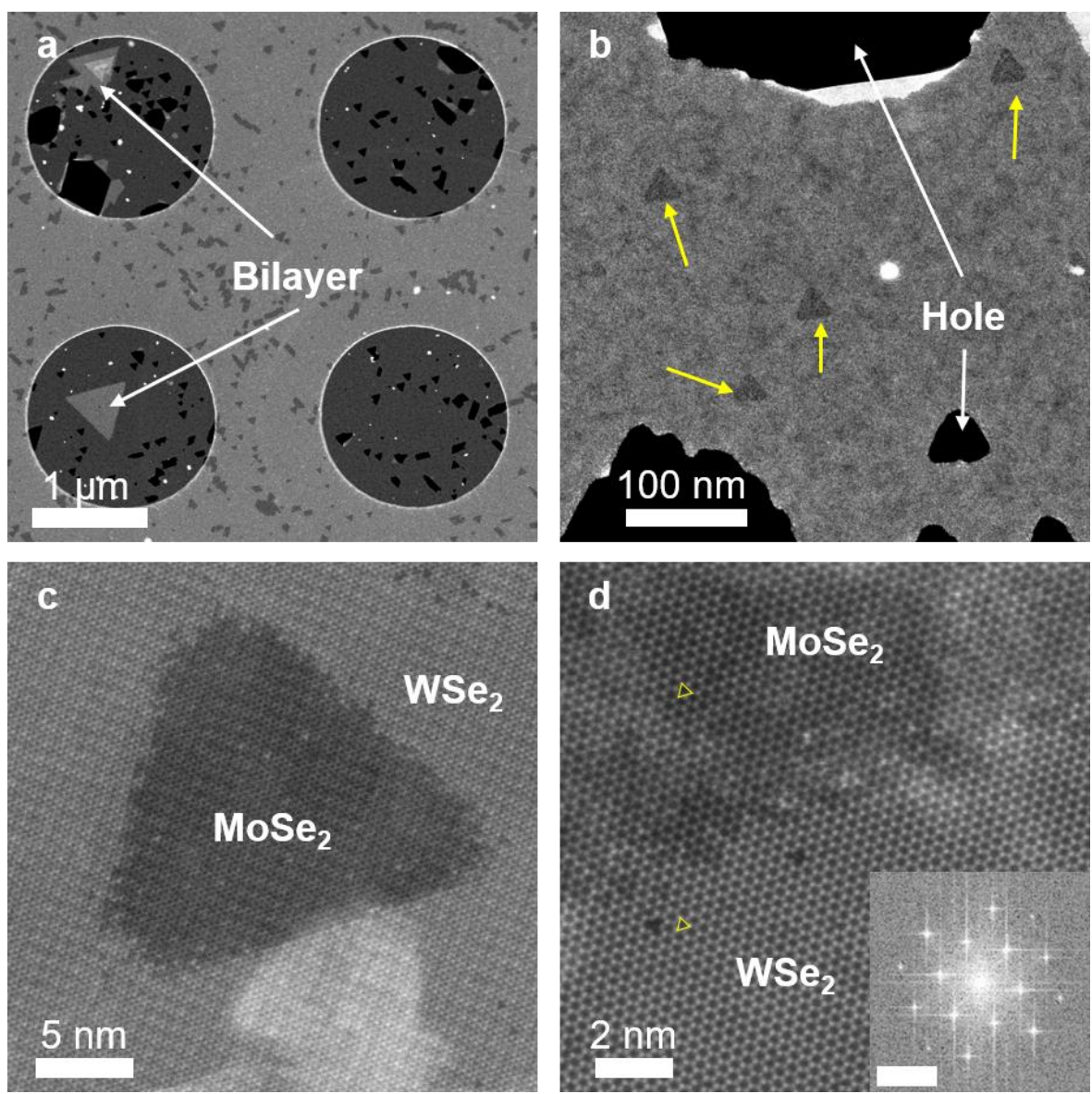

**Figure S1. ADF-STEM imaging of the heterostructure.** (a) Low-magnification ADF-STEM image of the MoSe$_2$/WSe$_2$ nanodot/matrix in-plane heterostructure showing an almost fully coalesced monolayer film with occasional WSe$_2$ bilayers on top; the triangular and trapezoidal holes seen in the film are the uncoalesced areas of the heterostructure, (b) higher magnification image confirming the presence of MoSe$_2$ dots embedded in the matrix of WSe$_2$, owing to the atomic number difference between the Mo and W; holes in the monolayer are also seen; contrast between the nanodots and the matrix is blurred by the presence surface carbon webbing, (c) atomic-resolution image highlighting the truncated triangular shape of a MoSe$_2$ triangle (the image is obtained from a heterostructure film that is encapsulated with hBN flakes on top and bottom and thus shows moiré fringe contrast); the white hazy contrast in the bottom part of the image is from the interfacial carbon webbing present between the layers in the stack, and (d) higher magnification image showing the atomically sharp interface minimal alloying; matching orientations of metal atom positions (yellow triangles) in the image, and FFT spots in the **inset** reveal the lateral epitaxy between the dot and the matrix (scale bar in the inset is 5 nm$^{-1}$).

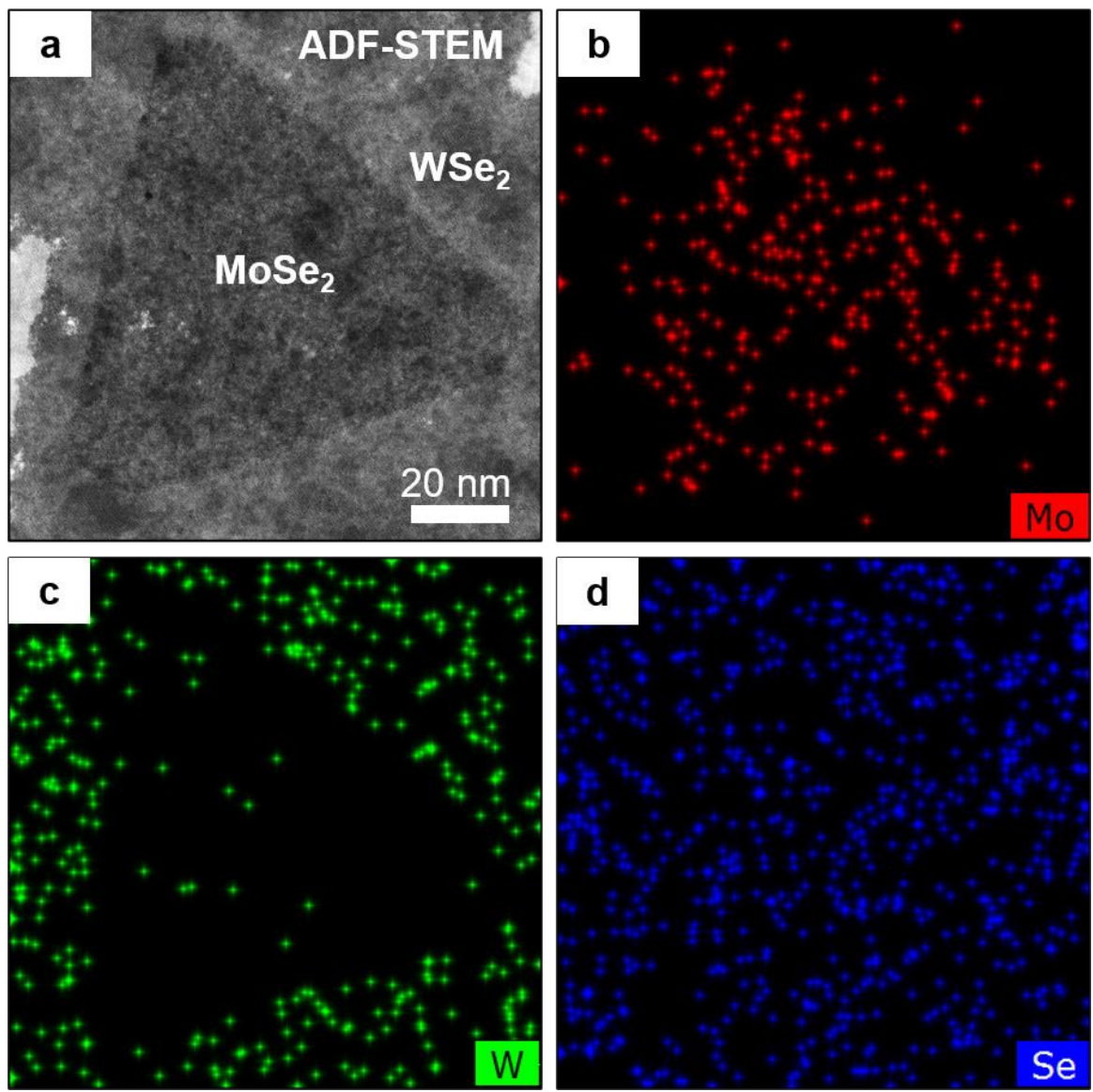

**Figure S2. STEM-EDS elemental mapping.** (a) ADF-STEM image of a MoSe₂ nanodot surrounded by WSe₂ matrix; contrast between the nanodot and the matrix is blurred by the presence surface carbon webbing, and (b–d) STEM-EDS elemental maps of Mo, W and Se showing the clear separation of metal signals between the nanodot and the matrix.

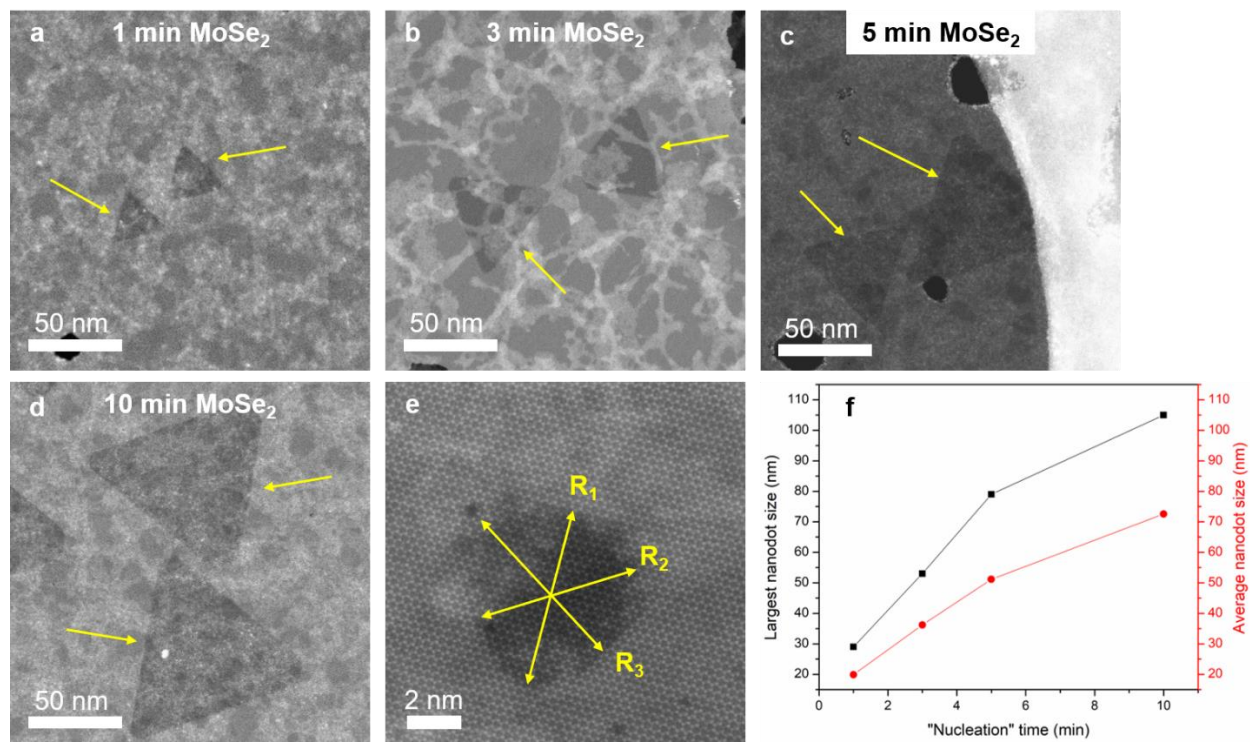

**Figure S3. MoSe₂ nanodot size vs nucleation time.** (a–d) ADF-STEM images of 1-, 3-, 5- and 10- min samples with similar field of view illustrating the increase in the dot size with the nucleation time; all the images show different amounts of surface carbon webbing on the heterostructure films which reduces the contrast between the nanodots and the matrix, (e) demonstration of the method used to measure the dot sizes using atomic-resolution image in **Figure 1a**; we measured the three distances between the truncated corners and the opposite sides of the triangles (R₁, R₂ and R₃) and then averaged them to get the triangle size, and (f) variation of the largest and average nanodot sizes observed among the datasets investigated in this study against the nucleation time.

It is also clear from the scatter plot in **Figure 1e** that each of the four samples contains dots with varying sizes, probably because a dot could nucleate at any moment during the nucleation step. So, for a given sample, it is more appropriate to define the largest possible nanodot size in a given sample rather than an average nanodot size. As seen from the plot, the largest nanodot sizes observed in the 1-, 3-, 5- and 10- min samples are 29, 53, 79 and 105 nm, respectively. When the largest and average dot sizes (among the dots investigated in this study) are plotted against the nucleation time (**Figure S3f**), we observe that they increase linearly with the time, at first, but slow down beyond five minutes.

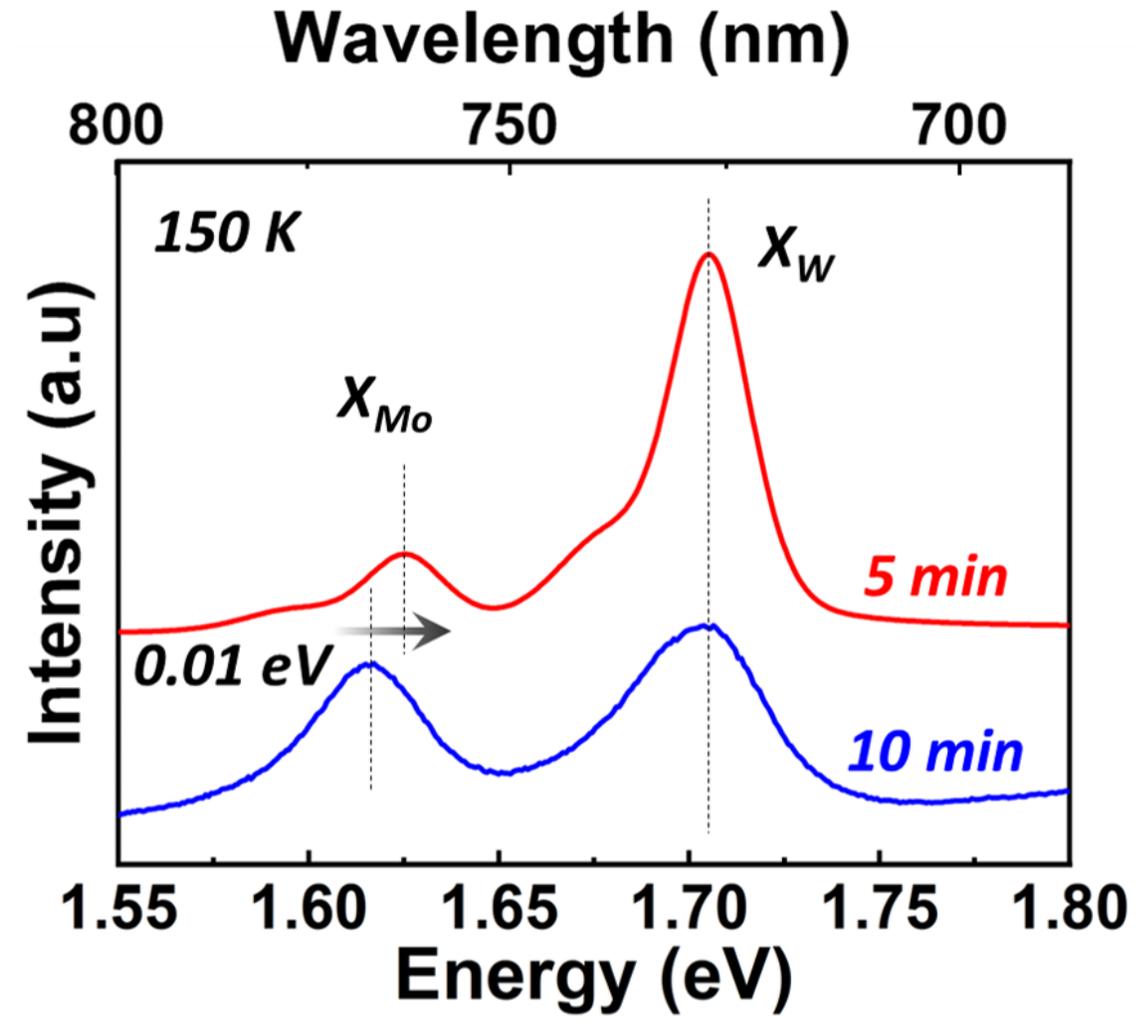

**Figure S4. PL spectra of 10-min (blue) and 5-min (blue) heterostructure samples obtained at 150 K.**

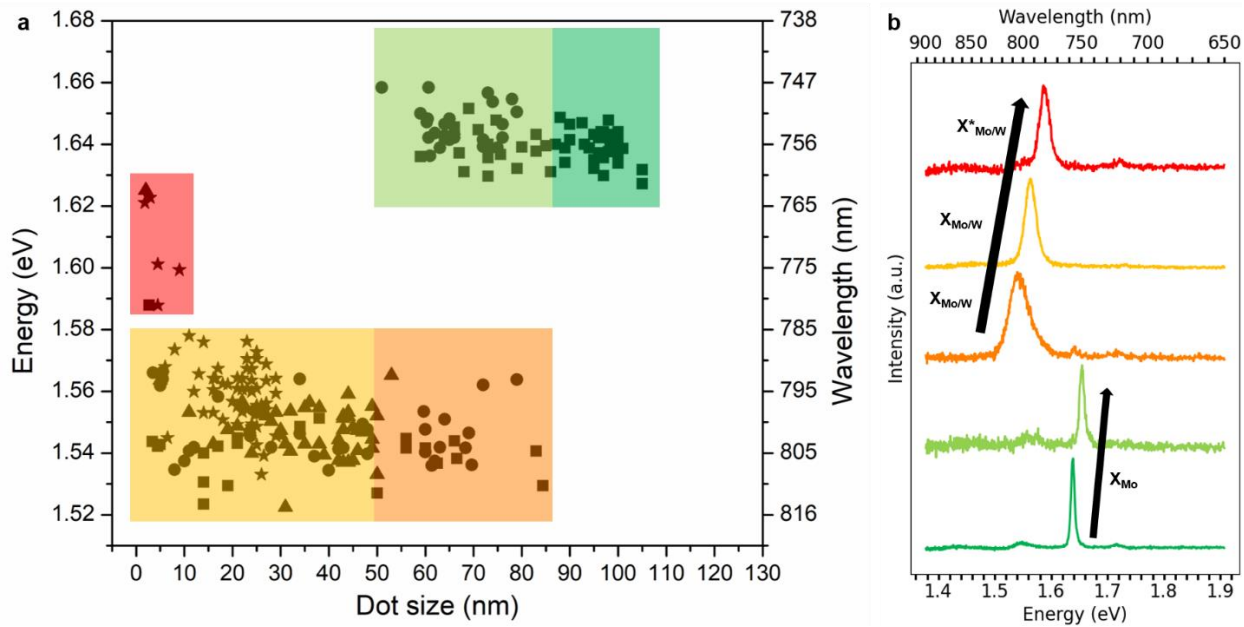

**Figure S5. Representative CL spectra from different regions of the scatter plot in Figure 1e.** (a) scatter plot shown in **Figure 1e** overlaid with five boxes shaded with different colors to represent five unique type of spectra we observed as a function of nanodot size, and (b) example CL spectra from five areas marked in (a) with matching colors.

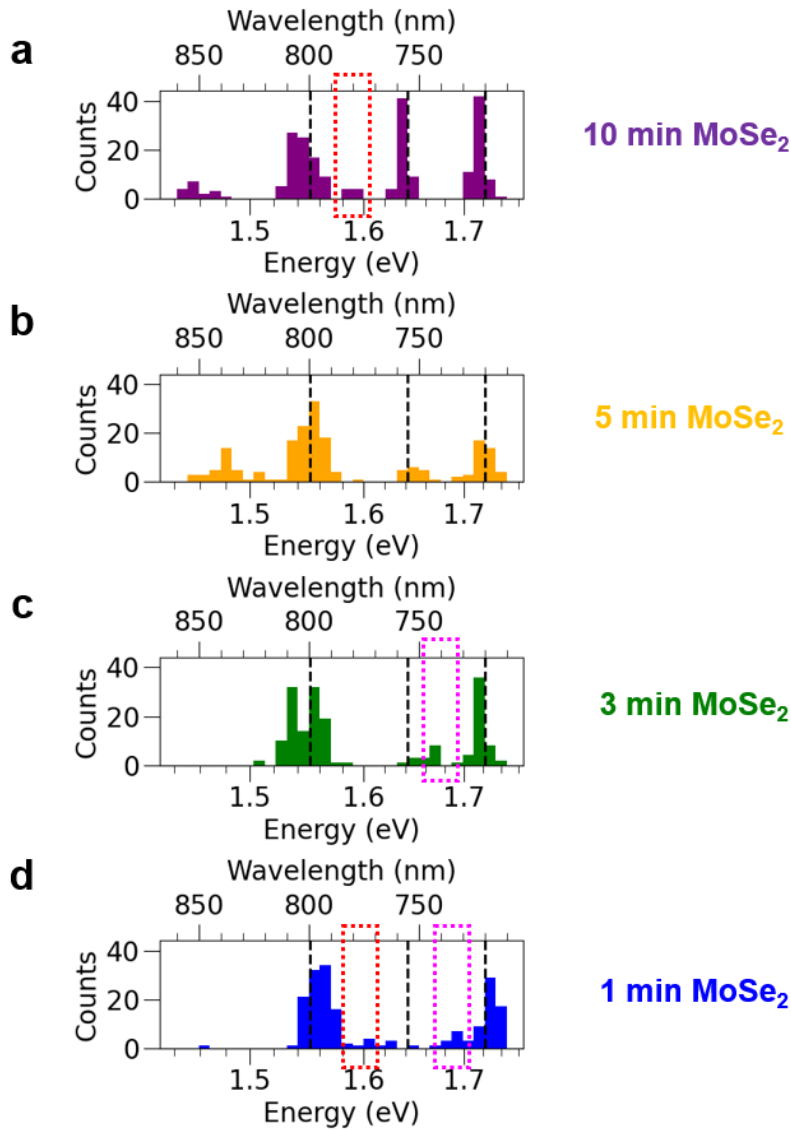

**Figure S6. Histograms of CL peak positions plotted separately for the four samples.** (a–d) Histograms showing the peak positions measured from different regions of the four samples with different MoSe$_2$ nucleation times: (a) 10-, (b) 5-, (c) 3- and (d) 1- min; the dashed black lines at 1.551 eV, 1.642 eV and 1.722 eV in the histograms represent the average wavelength values of X$_{Mo/W}$, X$_{Mo}$ and X$_W$ emissions, respectively; dashed red rectangles represent the X*$_{Mo/W}$ peaks and dashed magenta rectangles belong to WSe$_2$ trions (X$^T_W$).

The histograms showing the peak positions measured from different regions of the four samples are presented individually in **Figure S6a–d** (see **Methods** for the details about the generation of histograms). While all four samples exhibit the X$_W$ emission peaks at around 1.722 eV (used as the X$_W$ reference dashed black line in the histograms), there are subtle changes in the relative counts of the $X_{Mo}$ and $X_{Mo/w}$ peaks between the four samples. First, the $X_{Mo}$ peaks around 1.642 eV are predominantly present in the histograms of 10- and 5- min samples (**Figure S6a, b**) because longer nucleation time leads to the growth of $MoSe_2$ nanodots above 50 – 85 nm. Whereas, the $X_{Mo/w}$ peaks around 1.551 eV are present in all the four samples because a dot can nucleate at any point during the nucleation time and grow to intermediate size range (10 – 50 nm), suitable for the $X_{Mo/w}$ to dominate the light emission. Among the datasets collected in this study, we found seven nanodots that exhibit $X^*_{Mo/w}$ emission with wavelengths between 1.59 eV and 1.63 eV. Their peak positions are highlighted with a dashed red box in the histograms of 10- and 1- min samples (**Figure S6a, d**). We also observed less intense $WSe_2$ trion ($X^T_w$) emission peaks in 3- and 1- min sample at around 1.68 – 1.70 eV, as highlighted by magenta boxes in their histograms (**Figure S6c, d**).[1] We speculate that the broad distribution of peaks in the 1.38 – 1.51 eV range stem from midgap states induced by photoactive defects in the heterostructure layer.[2] Their unusually high count in the 5 min sample is believed to be due to its shorter growth step which likely resulted in more uncoalesced edges and defects in the film (see **Methods**).

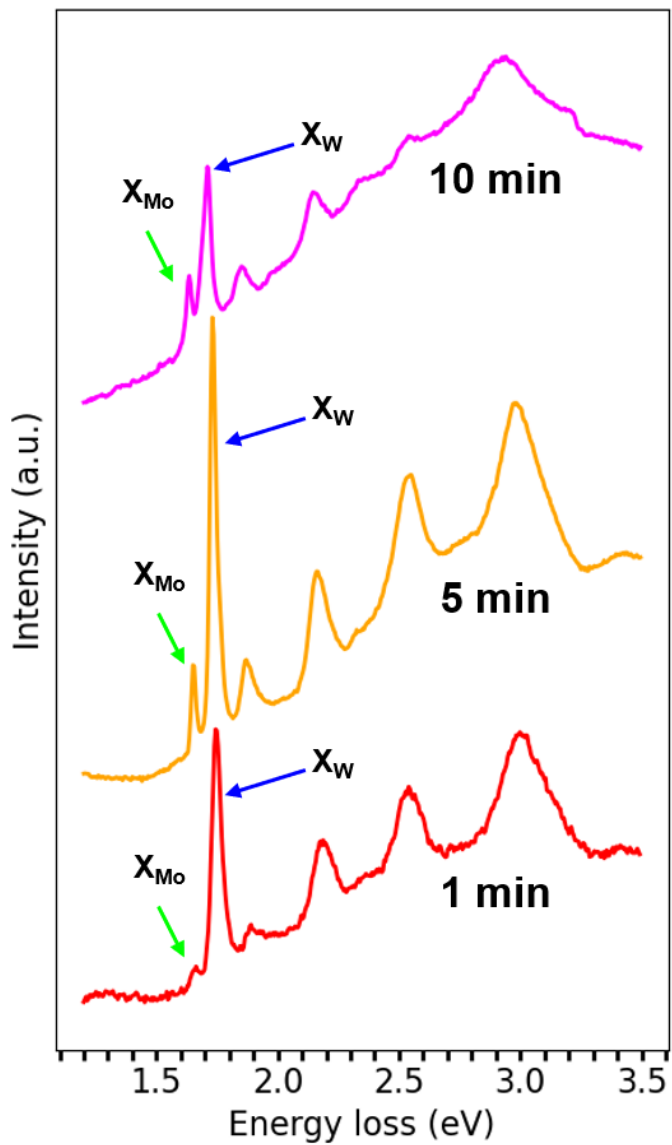

**Figure S7. EELS spectra obtained from 1-, 5- and 10- min samples.** All of them show A and B exciton peaks of $MoSe_2$ and $WSe_2$ and the relative intensity of the $MoSe_2$ A exciton peak increases with increasing nucleation time resulting from increasing area fraction of $MoSe_2$ with increasing dot size.

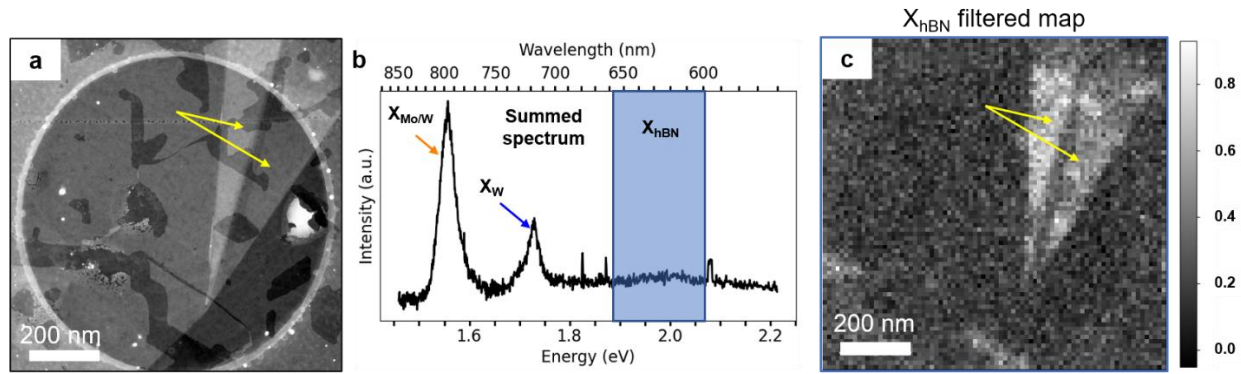

**Figure S8. Emission from hBN ($X_{hBN}$).** (a) low-magnification ADF-STEM image of the heterostructure film where the hBN is folded (yellow arrows), (b) summed CL spectrum obtained from the area shown in (a), and (c) filtered map of the broad $X_{hBN}$ peak illuminating the folds in the hBN.

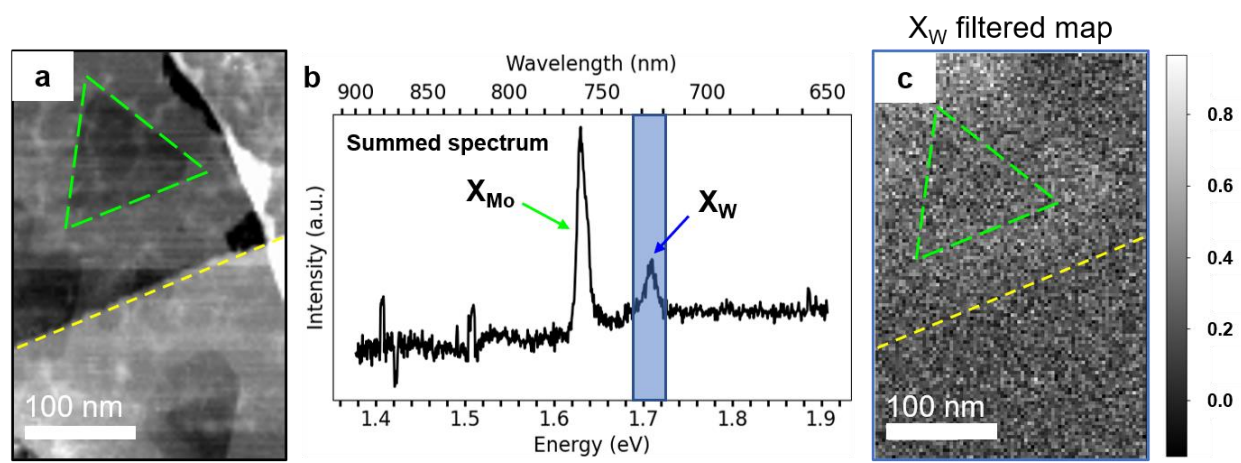

**Figure S9. Emission map of $X_W$.** (a) ADF-STEM image same as **Figure 3b**, (b) summed CL spectrum obtained from the area shown in (a), same as **Figure 3c**, and (c) filtered map of $X_W$ peak showing the emission around the top nanodot in the monolayer area.

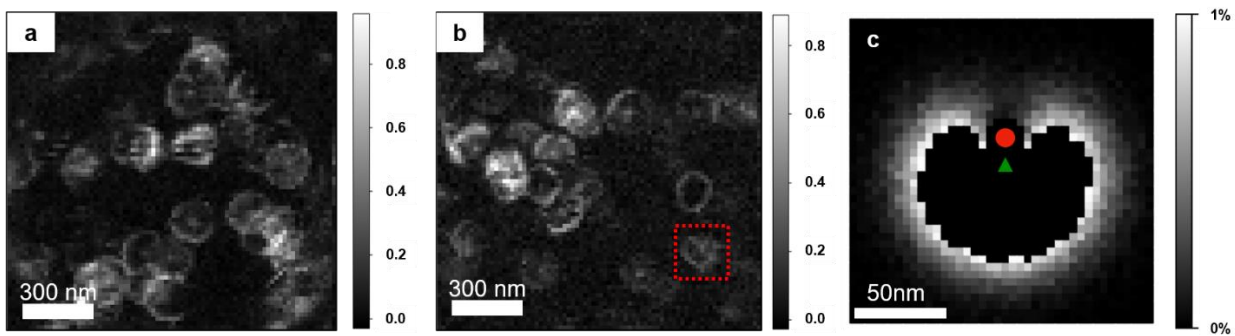

**Figure S10. Large-area emission maps of $X_{Mo/W}$ showing multiple ring patterns.** (a, b) Low-magnification filtered maps of $X_{Mo/W}$ peak showing multiple ring-pattern emissions; dashed red

box highlights a ring pattern that got distorted into a heart shape, and (c) demonstration of heart-shaped emission pattern formation using two nearby dots by extending the 2D random walk model from **Figure 4**; red dot represents the second dot.

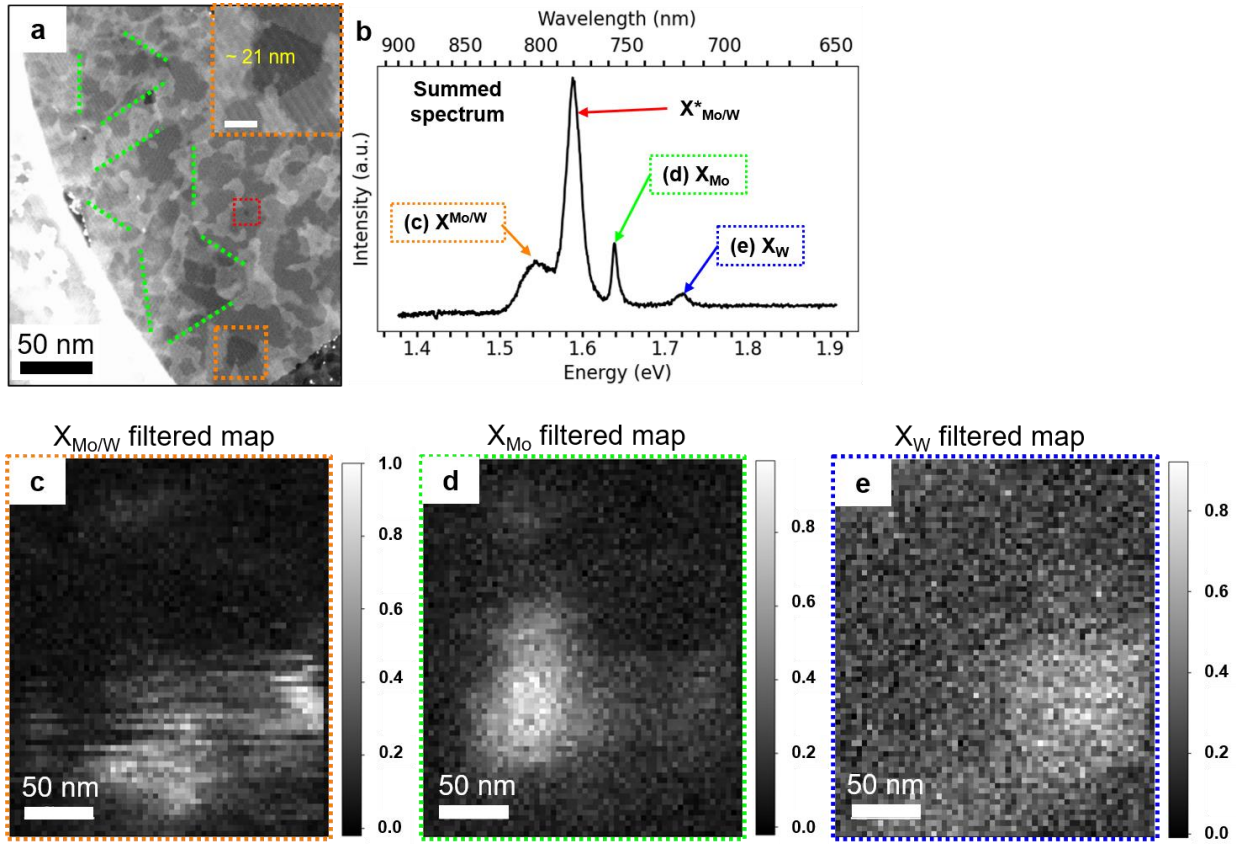

**Figure S11. Emission maps of the other peaks seen in Figure 5b.** (a) ADF-STEM image same as **Figure 5a**; large dots emitting the $X_{Mo}$ peak are outlined with dotted green lines; dot marked with orange box is enlarged and shown in the inset and is ~ 21 nm in size, (b) summed spectrum obtained from the area shown in (a), same as **Figure 5b**, (c) filtered map of $X_{Mo/W}$ peak showing the signal is delocalized from interaction with neighboring dots, (d) filtered map of $X_{Mo}$ peak showing the emission confined within the large nanodots outlined by dotted green lines, and (e) filtered map of $X_W$ peak. Scale bar in the panel (a) inset is 10 nm.

The 21 nm sized dot (outlined by dotted orange square and magnified in **Figure S11a** inset) emits light at 1.545 eV, attributed to the $X_{Mo/W}$ emission, and its emission map appears delocalized (**Figure S11c**) from interaction with neighboring dots. The largest nanodots in **Figure S11a** are outlined by dotted green lines wherein two of them appear merged with each other in the shape of an hourglass. These nanodots emit light at 1.639 eV corresponding to the $X_{Mo}$ emission, as revealed by its filtered map in **Figure S11d**. The peak at 1.723 eV in the summed CL spectrum (**Figure S11b**) corresponds to the $X_W$ emission and its filtered map is presented in **Figure S11e**.

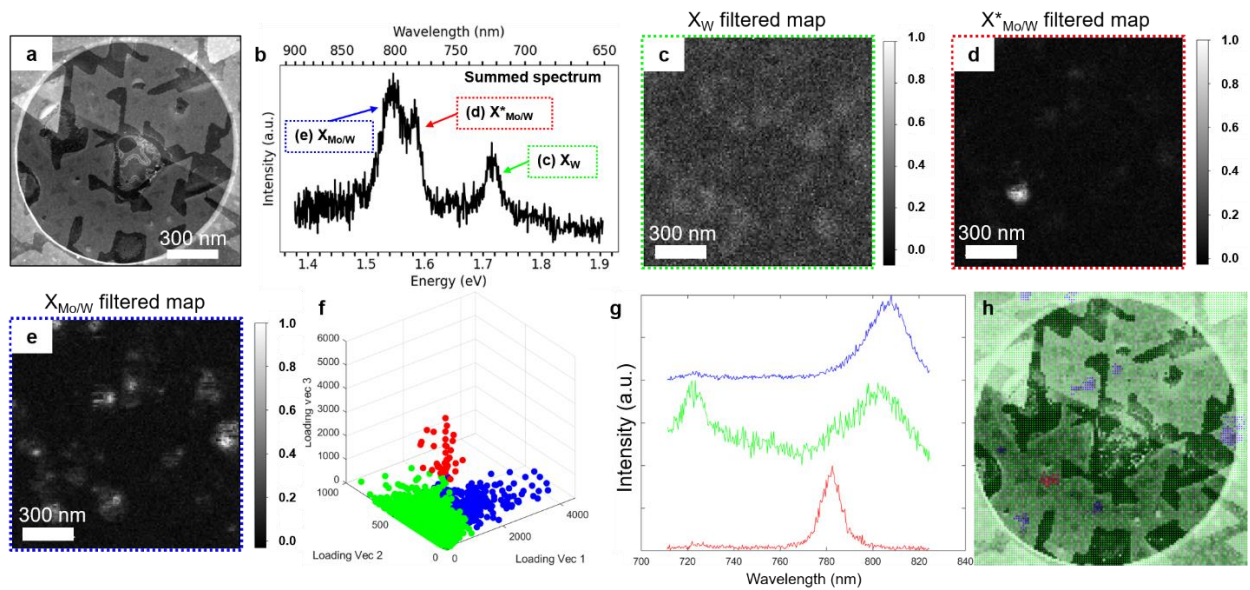

**Figure S12. Comparison between NMF+GMM and wavelength filtering.** (a) Low-magnification ADF-STEM image of heterostructure encapsulated in hBN, showing multiple $MoSe_2$ nanodots, (b) summed CL spectrum obtained from the area shown in (a) showing the $X_W$, $X^*_{Mo/W}$ and $X_{Mo/W}$ emissions, (c–e) filtered maps of $X_W$, $X^*_{Mo/W}$ and $X_{Mo/W}$ emissions, (f–h) NMF+GMM analysis performed on the same dataset: (f) scatter plot of the coefficients of the three NMF loading vectors divided into three clusters (green, red and blue), (g) class-average spectra of the three clusters, and (h) map of the three clusters observed in (f) and (g) overlaid onto the ADF-STEM image in (a), confirming that the positions of the three clusters match the three emission modes (green: $X_W$, red: $X^*_{Mo/W}$ and blue: $X_{Mo/W}$).